\documentclass[letterpaper, conference]{ieeeconf}

\IEEEoverridecommandlockouts                              % This command is only
\usepackage{multicol}
\usepackage{amsmath}
\usepackage{amsfonts}
\usepackage{amssymb}
\usepackage[dvips]{graphicx}
\usepackage{float}
\usepackage{subfigure}
\usepackage{psfrag}
\usepackage{balance}
\usepackage{multirow}
\usepackage{algorithm}
\usepackage{algpseudocode}

\usepackage{color}
\usepackage[colorlinks=false, pdfborder={0 0 0}]{hyperref}


\graphicspath{{figs/}}

\def\RR{\mathbb{R}}

\newcommand{\changefunction}[1]{%
	\expandafter\renewcommand\csname#1\endcsname[1][]%
	{\qopname\relax o{#1}\ifx\relax##1\relax\else^{##1}\fi}}
\changefunction{sin}
\changefunction{cos}
\renewcommand{\cos}{\,\text{c}}
\renewcommand{\sin}{\,\text{s}}

\begin{document}

	\title{Learning-based Parameter Optimization for a Class of Orbital Tracking Control Laws}
	
	\author{Gianni Bianchini$^1$, Andrea Garulli$^1$, Antonio Giannitrapani$^1$, Mirko Leomanni$^2$, Renato Quartullo$^1$\thanks{$^1$Dipartimento di Ingegneria dell'Informazione e Scienze Matematiche, Universit\`a di Siena, Italy; $^2$Dipartimento di Ingegneria, Universit\`a di Perugia, Italy. Email: {\{giannibi,garulli,giannitrapani,quartullo\}@diism.unisi.it; mirko.leomanni@unipg.it}.}%
	}
	\maketitle
	
	\begin{abstract}
		This paper presents a machine learning approach for tuning the parameters of a family of stabilizing controllers for orbital tracking.
		An augmented random search algorithm is deployed, which aims at minimizing a cost function combining convergence time and fuel consumption.
		The main feature of the proposed learning strategy is that closed-loop stability is always guaranteed during the exploration of the parameter space, {a property that allows one to streamline the training process by restricting the search domain to well-behaved control policies.} The proposed approach is tested on two case studies: an orbital transfer and a rendezvous and docking mission. It is shown that in both cases the learned control parameters lead to a significant improvement of the considered performance measure.
	\end{abstract} 
	
	\section{Introduction}
	Learning control laws from data has always been a primary objective of the control research community, leading to a vast body of results in areas such as adaptive control \cite{AWbook13,KKKbook95}, iterative learning control \cite{MDB92,BTA06}, direct control estimation \cite{CLS02,FN15} and reinforcement learning \cite{Ber19book,Rec19}. 
	Recently, this research field has seen a renewed interest due to the impressive progress of reinforcement learning techniques \cite{MKSR15,SHMG16}.
	The application of such techniques to continuous control problems have indeed proven to be successful in addressing complex tasks \cite{LHPH15}, with specific contributions in different areas including robotic manipulation \cite{KIPI18,ABCJ20}, mobile robotics \cite{GBKD21}, locomotion \cite{PBYV17}, power systems \cite{ZZQ19} and many others.
	
	Learning-based control approaches have recently been applied to guidance and control problems in the aerospace field, see, e.g., \cite{izzo2019survey,shirobokov2021}.
	In particular, rendezvous and docking (RVD) problems have been tackled by machine learning techniques in combination with model-based methods in \cite{Xia2016,Ueda2019,Hongjue2020}, as well as by reinforcement learning \cite{Gaudet2018,Wang2020,hovell2020deep}. %In particular, machine learning techniques have been used in combination with traditional model-based techniques in rendezvous and docking (RVD) problems \cite{Xia2016,Ueda2019,Hongjue2020}. Recently, there has been also an intense work on reinforcement learning approaches to such problems \cite{Gaudet2018,Wang2020,hovell2020deep}.
	These methods allow one to optimize a variety of performance indexes, related to different aspects of the mission to be accomplished. 
	
	The widespread application of learning-based approaches has also raised a number of fundamental questions related to issues such as stability, performance guarantees, and robustness. 
	In fact, safety-critical applications like those involving RVD maneuvers require special care to ensure robust stability and performance of the designed guidance and control schemes.
	In this respect, two alternative approaches can be taken. One is the recent research effort towards results that guarantee stability of feedback schemes including neural controllers (see e.g. \cite{BTSK17,YSA21,WBRM21,newton2022}).
	The other main line of research exploits machine learning tools to learn controllers belonging to pre-specified families, whose structure is designed in order to guarantee the desired properties. Among the large number of contributions, \cite{RMT11} is one of the first works enforcing specific parameterizations of the controller (including the Youla-Kucera one) and learning its parameters using the REINFORCE algorithm \cite{Wil92}. The Youla parameterization is also adopted in \cite{FB17}, while PID controllers are considered in \cite{LSLF20,LFLM22}. Learning within a family of robustly stabilizing controllers has been addressed in \cite{HST20}.
	
	Herein, we leverage the latter approach, to address the problem of learning the parameters of a control law for orbital tracking.
	A major source of complexity for such type of nonlinear control problems is the requirement to ensure closed-loop stability while minimizing a nonsmooth performance index describing a trade-off between fuel consumption and maneuver completion time. To address this challenge, in this paper we restrict our attention to the family of nonlinear stabilizing controllers introduced in \cite{leomanni2017class} and propose a learning strategy for tuning the controller parameters. The learning algorithm can be seen as a specialized version of REINFORCE and requires only the computation of the cost value associated to a simulation of the closed-loop control system (episode). The main benefit of the proposed approach is that closed-loop stability is always guaranteed for each episode during the exploration of the parameter space. Besides ensuring that the learned control policy is stabilizing, this feature also significantly speeds up the learning process. Simulated case studies show that the resulting controller provides the desired trade-off between settling time and fuel consumption, in two relevant RVD missions.
	
	The paper is organized as follows. Section \ref{sec:pf} reviews the orbital tracking model, and introduces the class of stabilizing controllers along with the associated optimal control problem. The learning algorithm is presented in Section \ref{sec:la}. The case studies concerning an orbital transfer and a rendezvous mission are reported and discussed in Section \ref{sec:ns}, while Section \ref{sec:conc} contains conclusions and future developments.
	
	\subsection*{Notation}
	{
		$\RR^n$ is the real $n-$space, and $\mathbb{Z}$ denotes the set of integer numbers; for a real vector or matrix $x$, $x^T$ denotes its transpose and $||x||$ its Euclidean norm.  The symbol ${0}_{n\times m}$ denotes a null $n\times m$ matrix, while the identity matrix of order $n$ is denoted by ${I}_n$. The partial derivative ${\partial f}/{\partial x}$ is expressed as a row vector. To save space, ${\rm cos}(\cdot)$ and ${\rm sin}(\cdot)$ are abbreviated with $\text{c}(\cdot)$ and $\text{s}(\cdot)$, respectively. Moreover, we define the rotation matrix
		\vspace{-1mm}
		\begin{equation*}
			R(\phi)=
			\left[
			\begin{array}{l r}
				\text{c}(\phi) & -\text{s}(\phi)\\
				\text{s}(\phi) & \text{c}(\phi)
			\end{array}
			\right].
			\vspace{-1mm}
		\end{equation*}
		%	is the counter-clockwise rotation operator, by an angle $\phi$, in $\RR^2$.% The definition of the signum function follows the convention \vspace{-1mm}
		%	$$
		%	\text{sgn}(x)=\left\{
		%	\begin{array}{r l}
		%		1 & x>0\\
		%		0 & x=0\\
		%		-1 & x<0\,,
		%	\end{array}
		%	\right.\vspace{-1mm}
		%	$$
		%	where $x\in\RR$. The continuous time index is denoted by $t\in\RR^+$.
	}
	
	\section{Problem Formulation}
	\label{sec:pf}
	
	In this paper, the dynamics of an orbiting spacecraft are described in terms of the six Equinoctial Orbital Element $\psi = \left[\psi_1\,\ldots \, \psi_6\right]^T = \left[L,\,p,\,e_X,\,e_Y,\,h_X,\,h_Y\right]^T$, where $L$  is the true longitude, $p$ is the orbit semi-parameter, $e_X$, $e_Y$ are the components of the eccentricity vector, and $h_X$, $h_Y$ are the components of the inclination vector  \cite{walker1985set}.
	%Despite a little loss of physical meaning, equinoctial elements often are preferred to classical ones because they avoid singularity for circular and equatorial orbits, in which the argument of perigee and the longitude of ascending node are indeterminate, respectively. In this paper we restrict our attention to the case of closed Keplerian orbit (not parabolic), thus $\psi_3^2+ \psi_4^2 < 1$.
	The dynamics are given by
	\begin{equation*}\label{oediff}
		\dot{\psi} = f(\psi) + g(\psi)u,
	\end{equation*}
	% and the vector fields $f(\psi)$ and $g(\psi)$ are given by
	where $u = \left[u_{r},\,u_{\theta},\,u_{h}\right]^T$ is the control vector (radial, transverse and normal forcing accelerations, respectively),
	\begin{equation*}\label{fpsi}
		f(\psi)=\sqrt{\frac{\mu}{\psi_2^3}}\left[\begin{array}{cccccc} (1+\zeta_X)^2 & 0 & 0 & 0 & 0 & 0 \end{array} \right]^T,
	\end{equation*}
	\setlength\arraycolsep{0.4pt}
	\begin{equation*}
		g(\psi)=\frac{\sqrt{\psi_2}}{\sqrt{\mu}(1\!+\!\zeta_X)}\!\left[
		\begin{array}{ccc}
			0& 0 & {\eta} \\[1mm]
			0 &{2}\,\psi_2 &0 \\[1mm]
			\;\,({1\!+\!\zeta_X}) \sin(\psi_1) &  {q_X}&   - { \eta\, \psi_4}  \\[1mm]
			\!\!-({1\!+\!\zeta_X}) \cos(\psi_1) & {q_Y} &    \;\;\, {\eta\, \psi_3} \\[2mm]
			0& 0& \dfrac{(1\!+\!h^{\!2})}{2}\cos(\psi_1)  \\[2mm]
			0& 0 &\dfrac{(1\!+\!h^{\!2})}{2}\sin(\psi_1)
		\end{array}\right]\!,\label{gpsi}
	\end{equation*}
	%\begin{eqnarray*}
	%\left[
	%\begin{array}{l}
	$\zeta_X = \psi_3\cos(\psi_1)+\psi_4\sin(\psi_1)$,
	$q_X = \psi_3 +(2+\zeta_X) \cos(\psi_1)$,
	$q_Y = \psi_4 + (2+\zeta_X)\sin(\psi_1)$,
	$\eta = \psi_5\sin(\psi_1)-\psi_6\cos(\psi_1)$,
	$h^2=\psi_5^2+ \psi_6^2$,
	%\end{eqnarray*}
	and $\mu$ is the gravitational parameter of the central body.
	On any unforced orbit, only the true longitude $\psi_1$ varies in time.
	
	The  considered control task is to track a target reference trajectory
	$\psi^r(t)=[\psi_1^r(t),\psi_2^r,\psi_3^r,\psi_4^r,\psi_5^r,\psi_6^r]^T$
	where $\psi^r(t)$ satisfies the unforced periodic dynamics $\dot{\psi}^r= f(\psi^r)$
	%\begin{equation}\label{oediffr}
	%	\dot{\psi^r}= f(\psi^r)
	%\end{equation}
	with given initial conditions $\psi^r(0)$. %= [L^r(0), {p}^r, {e}_X^r, {e}_Y^r, {h}_X^r, {h}_Y^r]^T$.
	The dynamics of the tracking error $\tilde\psi=\psi-\psi^r$ are modeled as in \cite{leomanni2017class} using the transformed variables:
	%.
	%Then, the error dynamics evolves according to the time-varying system
	%\begin{equation}\label{errorsys}
	%	\dot{\tilde\psi}=\tilde{f}(\tilde\psi;\psi^r)+ g(\tilde\psi+\psi^r) u,
	%\end{equation}
	%where $\tilde{f}(\tilde\psi;\psi^r)=f(\tilde\psi+\psi^r)-f(\psi^r)$.
	%and define the following diffeomorphic coordinate transformation $x = x(\tilde{\psi},\psi^r)$:
	\begin{equation}\label{eq:change}
		\begin{array}{rcl}
			x_1 \;\,&=&\tilde\psi_1 \\ [2mm]
			x_2 \;\,&=&\sqrt{1+\frac{\tilde\psi_2}{\psi_2^r}}-1 \\[2mm]
			\left[\begin{array}{l}x_3\\x_4\end{array}\right]
			\!\!\!&~=~& \!\!
			\left[\begin{array}{c c}
				\frac{\psi_2^r}{\tilde\psi_2+\psi_2^r} & 0\\ 0 & \sqrt{\frac{\psi_2^r}{\tilde\psi_2+\psi_2^r}}
			\end{array}\right]\!R(\tilde\psi_1\!+\!\psi_1^r)\!\left[\begin{array}{r} \tilde\psi_3\!+\!\psi_3^r \\\!-\tilde\psi_4\!-\!\psi_4^r \end{array}\right] \\[5mm]
			&&\hspace{8mm}
			+\left[\begin{array}{c} -\frac{\tilde\psi_2}{\tilde\psi_2+\psi_2^r}   \\0\end{array}\right]- \left[
			\begin{array}{l}
				\zeta^r_X\\
				\zeta^r_Y
			\end{array}
			\right] \\
			x_5 \;\,&=&\tilde\psi_5 \\%[1mm]
			x_6 \;\,&=&\tilde\psi_6,
		\end{array}
	\end{equation}	
	where $[\zeta_X^r,\ \zeta_Y^r]^T=R(\psi_1^r)[\psi_3^r,\ -\psi_4^r]^T$.
	%%\begin{eqnarray*}
	%	\left[
	%	\begin{array}{l}
	%		\zeta_X^r\\
	%		\zeta_Y^r
	%	\end{array}
	%	\right]
	%	\!\! & = &
	%	R(\psi_1^r)
	%	\left[
	%	\begin{array}{r}
	%		\psi_3^r\\
	%		-\psi_4^r
	%	\end{array}
	%	\right].
	%%\end{eqnarray*}
	The transformation \eqref{eq:change} is such that $x=0$ if and only if $\tilde{\psi}=0$. The corresponding dynamic model is then given by:
	%The evolution of the tracking error, coming from the difference between dynamics described in \eqref{oediff} and \eqref{oediffr} for the state vector \eqref{ccx1}-\eqref{ccx6} reads
	\begin{equation}\label{oediff3}
		\dot{x}=
		\left[\begin{array}{c}
			F(\chi,\psi^r) \\
			0_{2\times1}
		\end{array}
		\right]
		+
		\left[\begin{array}{c}
			G (\chi,\psi^r)\\
			0_{2\times2}
		\end{array}\right]
		\left[\begin{array}{c}
			u_r\\
			u_\theta
		\end{array}\right]
		+
		H(x,\psi^r)\,
		u_h,
	\end{equation}
	where $\chi=[x_1 \dots x_4]^T$,
	$$	
	{F}(\chi,\psi^r)\!\!=\!\!
	\left[\begin{array}{cccc} 0 & F_{12} & F_{13} & 0 \\
		0 & 0 & 0 & 0 \\
		0 & 0 & -F_{33} & -F_{12} \\
		0 & F_{42} & F_{12}\!+\!F_{43} & 0
	\end{array}\right]\!\chi,~
	%$$
	{G}(\chi,\psi^r)\!\!=\!\! \left[
	\begin{array}{cc}
		0&0  \\
		0 & G_{22} \\
		0&0   \\
		G_{41} & 0
	\end{array}\right],%\label{facto2a} % \vspace{2mm}
	$$
	$$
	\begin{array}{l}	
		H(x,\psi^r)=\\[1mm]
		\dfrac{G_{22}}{(x_2+1)}
		\left[
		\begin{array}{c}
			(x_5+\psi_5^r)\sin(x_1+\psi_1^r)-(x_6+\psi_6^r)\cos(x_1+\psi_1^r)\\[4mm]
			0_{3\times1}\\[4mm]
			\dfrac{1+(x_5+\psi_5^r)^2+(x_6+\psi_6^r)^2}{2}\cos(x_1+\psi_1^r)\\[4mm]
			\dfrac{1+(x_5+\psi_5^r)^2+(x_6+\psi_6^r)^2}{2}\sin(x_1+\psi_1^r)
		\end{array}
		\right],
	\end{array}
	$$
	$$
	\begin{array}{c}
		F_{12}\!\!=\!\!\sqrt{\frac{\mu}{({\psi_2^r})^3}}\left(x_3+1+{\zeta}_X^r\right)^2,\\[3mm]
		F_{13}\!\!=\!\!\sqrt{\frac{\mu}{({\psi_2^r})^3}}\left(x_3+2+2{\zeta}_X^r\right),\\[3mm]
		F_{42}\!\!=\!\!\sqrt{\frac{\mu}{({\psi_2^r})^3}}\left(x_2+2\right)\left(x_3+1+{\zeta}_X^r\right)^3,
	\end{array}$$
	$ F_{33}\!\!=\!\!F_{13}\,\zeta^r_Y$,
	$ F_{43}\!\!=\!\!F_{13}\,\zeta^r_X$,
	$ G_{22}\!\!=\!\!\sqrt{\frac{\psi_2^r}{\mu}}\frac{1}{(x_3+1+\zeta_X^r)}$, and
	$ G_{41}\!\!=\!\!\sqrt{\frac{\psi_2^r}{\mu}}.$
	The above vector fields are periodically time-varying with the same period as the reference trajectory.
	
	In \cite{leomanni2017class}, a class of stabilizing control laws for system \eqref{oediff3} is proposed. In this paper, we consider a parametric family of controllers that falls within the class introduced in \cite{leomanni2017class}, given by
	\begin{equation}\label{eq:controller}
		\resizebox{\hsize}{!}{$
			\begin{array}{rcl}
				u_r(x,\psi^r\!;K)&=&\displaystyle{-\frac{1}{G_{41}} \left(F_{43}\, x_3- \dot{\xi}  \right) - K_4(x_4-\xi) }  \\[3mm]
				u_\theta(x,\psi^r\!;K)&=&-\displaystyle{K_1\frac{G_{41}F_{12}}{G_{22}}\sin(x_1)
					-\frac{F_{42}}{G_{22}}(x_4\!-\!\xi) -K_2 \frac{G_{41}}{G_{22}}  x_2 }\\[3mm]
				u_h(x,\psi^r\!;K)&=&\displaystyle{- K_5 \frac{1}{G_{41}}\frac{\partial V}{\partial x} H }, %\!\left( \frac{\partial V(x,\psi^r)}{\partial x} \,H(x,\psi^r)\right),
			\end{array}$}
	\end{equation}
	where  $K=[K_1,\dots,K_5]^T$ is a vector of constant parameters, 
	\begin{equation}\label{Lyap2}
		V(x)=K_1 G_{41}(1-\cos(x_1))+\frac{1}{2}({x}_2^2+{x}_3^2+(x_4-\xi)^2+x_5^2 + x_6^2),
	\end{equation}
	and
	\begin{equation*}
		\xi=\frac{1}{F_{12}}\left(K_1 G_{41}{F_{13} \sin(x_1)}   -F_{33} x_3+{K_3 G_{41} x_3}\right).
	\end{equation*}
	The expressions of $\dot\xi$ and $\tfrac{\partial V}{\partial x} H$ in \eqref{eq:controller} are omitted for brevity. By using \eqref{Lyap2} as a Lyapunov function, it has been proved  that the control law \eqref{eq:controller} ensures almost global asymptotic stability of the origin of the closed-loop system \eqref{oediff3}, \eqref{eq:controller} for all $K_i>0,~i=1,\dots,5$ (see \cite{leomanni2017class} for details). However, assessing performance for such a design is not trivial.
	
	The goal of this paper is to tune the parameters $K$ of the stabilizing controller \eqref{eq:controller} so as to optimize the performance of the control system in terms of a trade-off between the settling time and the fuel consumption. To this purpose, let us denote by $y$ the distance between the actual and reference spacecraft position, expressed in Cartesian coordinates. This can be seen as an output signal of system \eqref{oediff3}, defined as
	\begin{equation}\label{eq:out}
		y=Y(x,\psi_r),
	\end{equation}
	where the mapping $Y$ is obtained from \eqref{eq:change} and the transformation which relates the satellite Equinoctial elements to the corresponding inertial cartesian states \cite{battin1999introduction}.
	System \eqref{oediff3}, \eqref{eq:out} with control law \eqref{eq:controller} is simulated over a horizon of length $T_e$ (each simulation is called an \emph{episode}). The input and output values collected at sampling times $kT_s$, $k=0,\dots,H$, with $T_e=H T_s$, are denoted as $u(k)$ and $y(k)$, respectively.
	The performance index to be minimized is then designed as
	\begin{equation}\label{eq:cost}
		J=H_{c} + \rho \sum_{k=0}^{H_{c}-1} ||u(k)||,
	\end{equation}
	where 
	
	\begin{equation}\label{eq:Hc}
	H_{c}= \min\{\bar{k}:~y(k) \leq \epsilon,~~\forall k \geq \bar{k}\},
\end{equation}
	is the number of samples necessary to achieve practical convergence and $\epsilon$ is a suitable threshold depending on the mission objective.
	The parameter $\rho$ is used to trade-off the two conflicting requirements of minimizing completion time and fuel. 
	%
	%
	%
	% Both the contributions are evaluated along a suitable sequence of time samples $\{kT_s\}_{k\in\mathbb{N}}$, where $T_s$ the sampling time. In particular, the settling time $T_{conv}$ is specified as the smallest $kT_s$ at which $V(x(kT_s))<\epsilon$, where $\epsilon$ is a threshold assessing practical convergence. Note that the set $\{x: \,V(x)\leq \epsilon\}$ is forward invariant. Moreover, the fuel consumption is measured as $\sum_{k=0}^{T_{conv}/T_s} ||u(kT_s)||$. The performance index $J$ to be minimized is then given by
	%\begin{equation}\label{eq:cost}
	%J=T_{conv} + \rho \sum_{k=0}^{T_{conv}/T_s} ||u(kT_s)||
	%\end{equation}
	%where the parameter $\rho$ is used to trade-off the two conflicting requirements of optimizing time and fuel.
	%{\color{red} Non si potrebbe formulare come un problema di massimizzazione andando a definire il reward? Nella versione precedente, la sommatoria andava fino a $T$ (lunghezza episodio), perchè? Perchè il funzionale di costo è espresso in funzione di $x(0)$ (dovrebbe dipendere anche dal tempo)?}
	%
	%
	%{\color{red} Spiegare qui sotto gli episodi e tutte le altre considerazioni che ho rimosso, verificare la compatibilità della nuova notazione per i samples}
	
	In order to optimize \eqref{eq:cost} with respect to the controller parameter vector $K$, a learning-based approach is pursued, as detailed in the next section.
	
	\section{Learning algorithm}
	\label{sec:la}
	
	Learning the parameter vector $K\in \mathbb{R}^q$ that optimizes the cost $J$ in \eqref{eq:cost} can be cast as a random exploration of the parameter space.
	A classical approach is the so-called \emph{random search}, which is based on a finite difference approximation of the function along a random search direction $\delta\in\mathbb{R}^q$, i.e.
	\begin{equation}
		\frac{J(K^{(i)}+\sigma\delta)-J(K^{(i)}-\sigma\delta)}{\sigma}
		\label{eq:findiff}
	\end{equation}
	where $J(K^{(i)})$ denotes the value of the cost \eqref{eq:cost} in an episode with parameter vector $K^{(i)}$, and $\sigma$ is a positive constant.  
	Then, the parameter vector $K$ is updated by taking a step along the direction $\delta$, proportional to the finite difference \eqref{eq:findiff}.
	
	An improved version of this approach is the Augmented Random Search (ARS) algorithm proposed in \cite{MGR18}, in which multiple random search directions $\delta_j \in\RR^q$, $j=1,\dots,N$, are selected in order to enhance the exploration of the parameter space. {In this paper, vectors $\delta_j$ are drawn from a normal distribution with zero mean and covariance matrix $\Sigma_{\delta}$.} For each direction, two perturbed parameter vectors $K_+^{(j)}$, $K_-^{(j)}$, $j=1,\dots,N$ are generated (in opposite directions). Then, the system is simulated for $2N$ episodes, one for each perturbed parameter, and the corresponding costs $J_+^{(j)}$, $J_-^{(j)}$, $j=1,\dots,N$, are computed according to \eqref{eq:cost}.
	Finally, the parameter vector is updated along a direction which is a weighted average of the random search vectors, according to the cost variation along each $\delta_j$. The update step is scaled by the standard deviation $\sigma_J$ of the cost values associated to the $2N$ episodes.
	The entire procedure is summarized in Algorithm \ref{alg:ars}. The outcome of the learning procedure is the value of the parameter vector at the last iteration, marked as $K^* = K^{(M)}$.
	
	\begin{algorithm}[t]
		\caption{Augmented Random Search (ARS)}\label{alg:ars}
		\begin{algorithmic}[1]
			\State Hyperparameters: number $M$ of iterations, stepsize $\alpha$, number $N$ of sampled directions per iteration, perturbation step $\sigma$, maximum length $H$ of each episode, covariance matrix $\Sigma_{\delta}$ of standard distribution for sampling vectors $\delta_j$.
			%covariance matrix of sampled directions $\Sigma$
			\State Initialize: parameter vector $K^{(1)} \in \RR^q$.
			\For {$i=1,2,\dots,M$}
			\State Sample independent vectors $\delta_j \in \RR^q$, $j=1,\dots,N$, from normal distribution with zero mean and covariance matrix $\Sigma_{\delta}$
			\For {$j=1,\dots,N$}
			\State Define perturbed parameter vectors
			$$\begin{array}{rcl}
				K_+^{(j)} &=& K^{(i)} + \sigma \delta_j \vspace*{1mm}\\
				K_-^{(j)} &=& K^{(i)} - \sigma \delta_j
			\end{array} 
			$$
			\State Simulate system with parameters $K_+^{(j)}$, $K_-^{(j)}$
			\State For each simulation, compute costs $J_+^{(j)}$, $J_-^{(j)}$
			\EndFor
			\State Compute standard deviation $\sigma_J$ of the $2N$ cost values $J_+^{(j)}$, $J_-^{(j)}$, $j=1,\dots,N$
			\State Update control parameters as
			$$K^{(i+1)}=K^{(i)} - \frac{\alpha}{N \sigma_J} \sum_{j=1}^N (J_+^{(j)} - J_-^{(j)})  \delta_j$$
			\EndFor
		\end{algorithmic}
	\end{algorithm}

	\section{Numerical simulations}
	\label{sec:ns}
	To demonstrate the benefit of the proposed methodology, in this section, Algorithm~\ref{alg:ars} is employed for tuning the parameter vector $K=\left[K_1,\ldots,K_5\right]^T$ of the control law proposed in Section \ref{sec:pf}, for two different case-studies. In particular, an orbital transfer from a geostationary transfer orbit (GTO) to a geostationary Earth orbit (GEO), and a rendezvous mission performed in a low Earth orbit (LEO), are considered.
	
	Algorithm~\ref{alg:ars} is implemented in C++ on a 3.10 GHz CPU with 16 cores, exploiting parallel computing. 
	%The parallelization is applied to both episodic exploration directions and initial conditions to optimize the computational efficiency.
	 The hyperparameters utilized in the learning algorithm are reported for each scenario in the corresponding subsection.
	%Likewise, the computation times of the algorithm are provided for each scenario due to their dependence on specific hyperparameters chosen for that case, such as the number of iterations or the length of each episode. \mycomm{??? ultima frase non chiara}
	
	\subsection{Orbital transfer}
	In this case study, the objective is to steer the satellite from an initial GTO (semi-major axis $a = 24364$ km, eccentricity $e= 0.7306$,  inclination $i = 63$ deg, right ascension of the ascending node $\Omega= 75$ deg, argument of periapsis $\omega= 52$ deg, initial true longitude $L =\pi/6$) to a circular equatorial GEO (semi-major axis $a^r= 42165$ km).
	The sampling time is $T_s= 45$ min and the parameters in~\eqref{eq:cost}-\eqref{eq:Hc} are set to $\rho = 50$ and $\epsilon = 10$ km.  The hyperparameters of Algorithm~\ref{alg:ars} are chosen as follows: $M=2000$, $\alpha = 5\cdot 10^{-3}$, $\sigma = 2\cdot 10^{-3}$, $N=16$, $H = 1280$, corresponding to 40 orbital periods along the target orbit and $T_e = 957.4$ hours. The initial parameter vector is chosen as $K^{(1)} = \left[0.1,\,1,\,1,\,1,\,10\right]^T$ and the covariance matrix for the search directions is taken as $\Sigma_{\delta}= \text{diag}\{0.1,\,1,\,1,\,1,\,10\}$. 
	
\begin{figure}[h]
		\centering
		\includegraphics[width=\columnwidth]{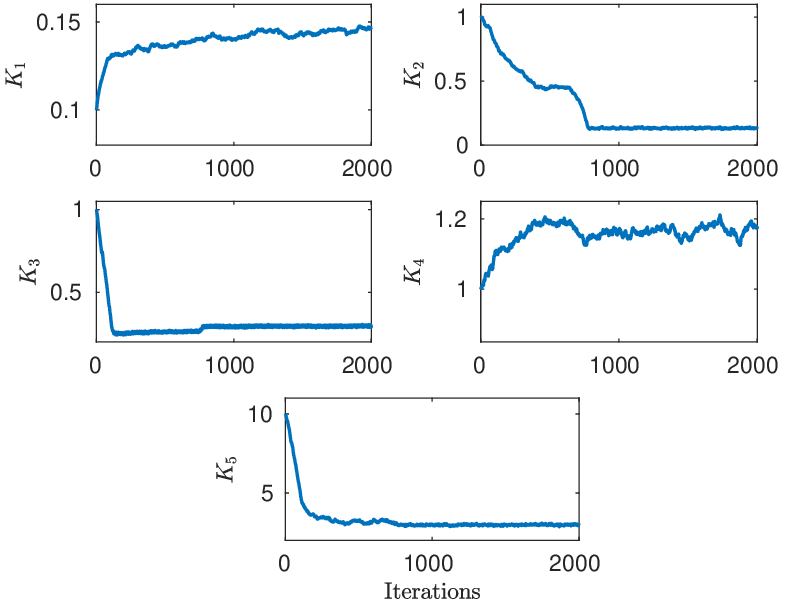}
		\caption{Scenario A. Evolution of the parameter vector $K^{(i)}$ during the learning phase.}\label{fig:K_OT}
	\end{figure}

\begin{figure}[h]
		\centering
		\includegraphics[width=\columnwidth]{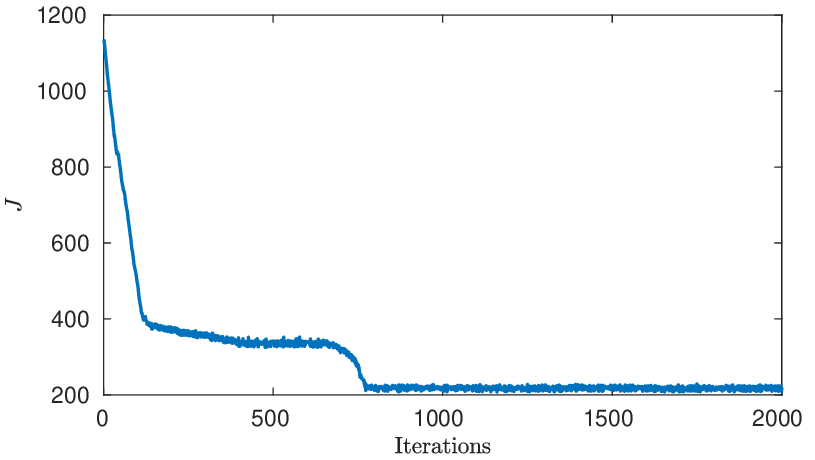}
		\caption{Scenario A. Evolution of the overall cost $J$.}\label{fig:cost_OT}
	\end{figure}

\begin{figure}[h]
		\centering
		\includegraphics[width=\columnwidth]{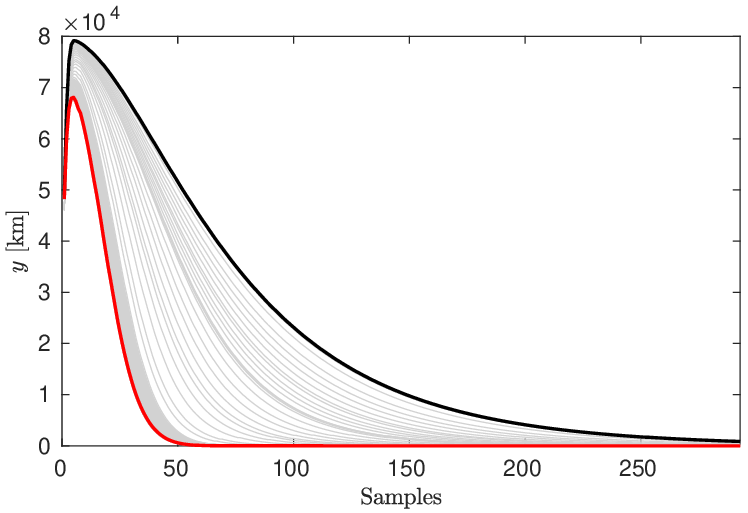}
		\caption{Scenario A. Distance $y$ generated by Algorithm~\ref{alg:ars}: initial (black line) and final iteration (red line).}\label{fig:dist_OT}
	\end{figure}

	The evolution of the parameter vector $K^{(i)}$ is depicted in Fig.~\ref{fig:K_OT}, while Fig.~\ref{fig:cost_OT} displays the corresponding cost $J$ defined by~\eqref{eq:cost}. The algorithm converges in less than 1000 iterations and leads to a cost reduction of about 82\% with respect to the initial maneuver cost, which is a remarkable improvement.	
	Fig.~\ref{fig:dist_OT} shows the distance $y$ in \eqref{eq:out} as a function of time, for all the iterations generated by the learning algorithm, where the black and red lines correspond to the first and last iteration, respectively. It can be noticed that optimizing the control parameters leads to a significant reduction of the flight time and that all the closed-loop system trajectories achieve asymptotic convergence, as expected.
	
	%In this case-study, the learning-based tuning is tested on a set of 50 different initial orbit conditions. The target orbit is an equatorial and circular GEO with semi-major axis of 42165 km, while the initial GTOs are characterized by a semi-major axis of 24364 km, eccentricity equal to 0.7306 and initial true longitude of $\pi/6$. The orientation of these orbits, determined by the inclination $i$, the right ascension of the ascending node (RAAN) $\Omega$ and the argument of periapsis $\omega$, is randomly chosen. In particular the Euler's angles $i$, $\Omega$ and $\omega$ are picked from a uniform distribution in the interval $\left[\pi/4,\,\pi/2\right]$. The algorithm hyperparameters are the same as in the previous case study, except for $T_s = 45$ min and $H = 1280$, corresponding to 40 target orbits and $T_e = 957.4$ hours. 
	
	{In order to further assess the effectiveness of the learning procedure, a more comprehensive analysis in which the satellite starts from 50 random initial conditions is performed.} 
	In particular, the chaser orbit shape is left unaltered, while the orientation angles $i$, $\omega$ and $\Omega$ are randomly drawn from a uniform distribution in the interval $\left[\pi/4,\,\pi/2\right]$. %This more extensive analysis aims to provide a more complete understanding of the learning procedure performance.
		
\begin{figure}[h]
		\centering
		\includegraphics[width=\columnwidth]{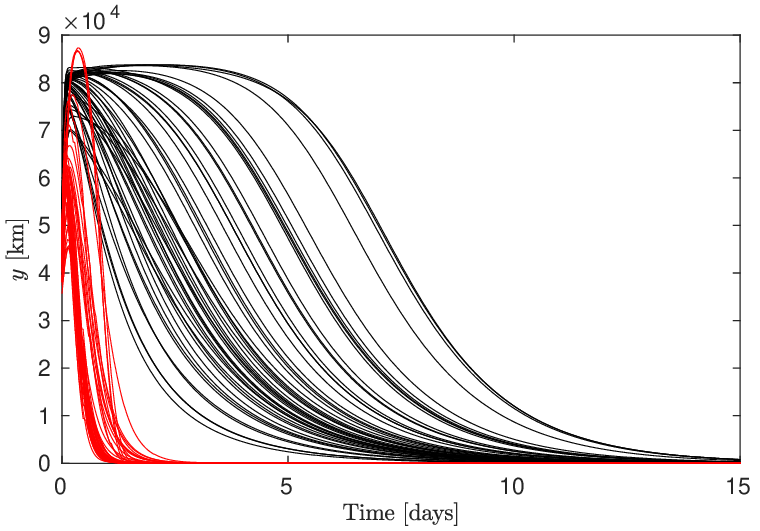}
		\caption{Scenario A, multiple transfers. All trajectories before (black) and after (red) learning procedure.}\label{fig:envelopTO}
	\end{figure}

	In Fig.~\ref{fig:envelopTO}, the trajectories optimized through Algorithm~\ref{alg:ars} (red) are compared with those corresponding to the initial choice of the controller parameters (black), for the considered set of initial conditions. It can be seen that the optimized trajectories achieve a much better convergence time as well as a smaller dispersion than the initial ones.
	The performance improvements in terms of total cost, convergence time, and fuel efficiency are reported in Table~\ref{tab:TOstat}. These results clearly show the effectiveness of the proposed learning technique in improving the performance of the closed-loop control system.
	%, which may lead to significant benefits in various real-world applications.
	
	\begin{table}[h]
		\centering
		\renewcommand{\arraystretch}{1.3}
		\begin{tabular}{|l|ccc|}
			\hline
			\multirow{2}{*}{} & \multicolumn{3}{c|}{Cost reduction (\%)} \\ \cline{2-4} 
			& \multicolumn{1}{c|}{Total cost} & \multicolumn{1}{c|}{Settling time} & Fuel consumption \\ \hline
			Average & \multicolumn{1}{c|}{78.4} & \multicolumn{1}{c|}{86.0} & 68.8 \\
			Minimum & \multicolumn{1}{c|}{65.4} & \multicolumn{1}{c|}{67.5} & 40.5 \\
			Maximum & \multicolumn{1}{c|}{83.6} & \multicolumn{1}{c|}{91.8} & 76.1 \\ \hline
		\end{tabular}
		\caption{Performance results for Scenario A with random initial conditions.}\label{tab:TOstat}
	\end{table}

	\subsection{Rendezvous}
	As a second case study, we consider a terminal rendezvous scenario, in which the target initially lies on a near-circular LEO  with an altitude of 1000 km above the Earth, inclination of 81 deg and initial true longitude of 45 deg. The chaser initial state is assumed to lie in the neighborhood of the target one. In practice, this situation arises after a preliminary orbit injection maneuver. A set of 50 initial conditions is generated through a normal distribution centered at the target equinoctial elements $\psi^r$ using the covariance matrix $\Sigma_{\psi}=\text{diag}\,\{0.5 \text{ deg},\,20 \text{ km},\,3\cdot10^{-5},\,3\cdot10^{-5},\,2\cdot10^{-3},\,2\cdot10^{-3}\}$. This corresponds to specifying an initial inter-satellite distance in the order of 60 km. 
	The sampling time is $T_s=3$ min, the parameters in~\eqref{eq:cost}-\eqref{eq:Hc} are set to $\rho = 400$ and $\epsilon = 1$ km and the number of iterations is $M = 5000$. The other hyperparameters are set equal to those of the previous case study. 
	
	The purpose of this setup is to assess the performance
	of a mean parameter vector $\hat{K}$ obtained by the learning
	process over all random initial conditions. This study is
	motivated by the fact that the initial condition of a terminal
	rendezvous mission is unknown beforehand and is
	the result of a previous transfer mission. In such a scenario,
	the controller tuning should ideally be performed on-board. However, the low amount of computational resources typically available on a spacecraft may preclude this possibility. To overcome this issue while still achieving an acceptable performance,
	pre-computing a mean parameter vector over a wide range
	of initial conditions turns out to be an effective alternative. A performance analysis of this approach is discussed in the
	following.
	
\begin{figure}[h]
		\centering
		\includegraphics[width=\columnwidth]{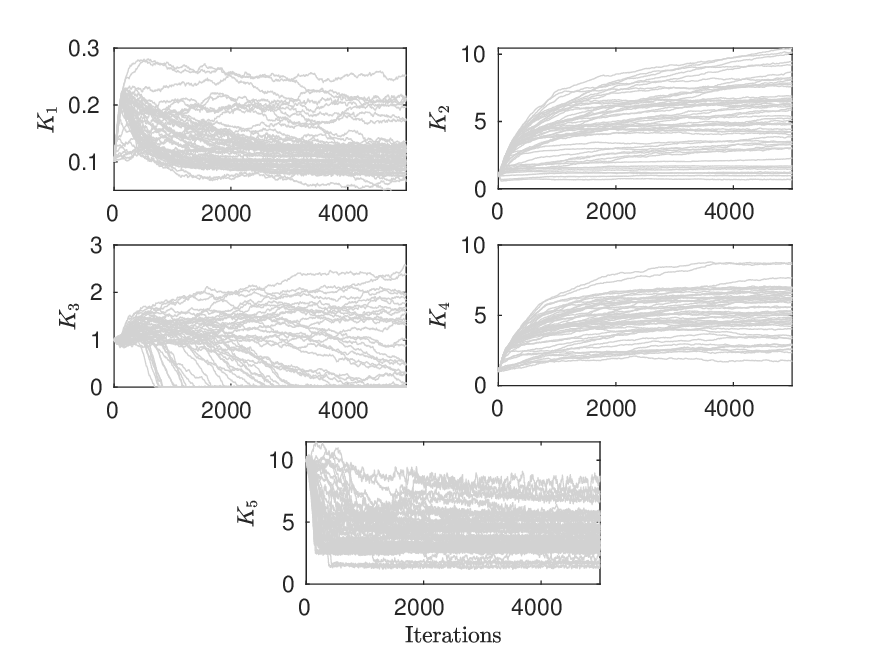}
		\caption{Scenario B. Parameter vectors during the learning for all the initial conditions.}\label{fig:K_RV}
	\end{figure}

\begin{figure}[h]
		\centering
		\includegraphics[width=\columnwidth]{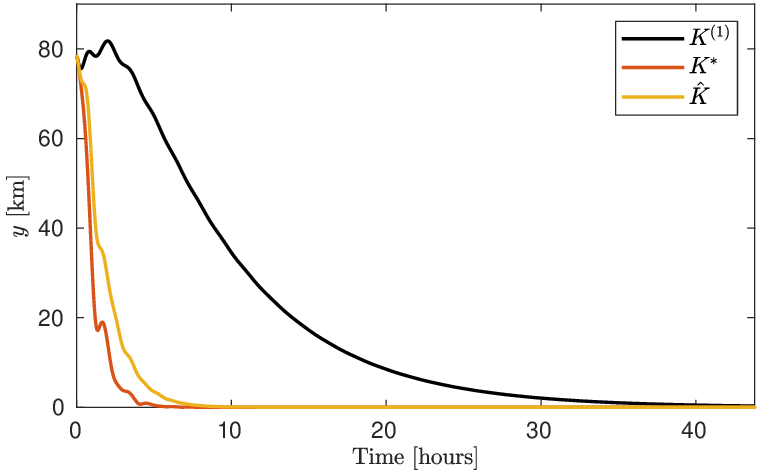}
		\caption{Scenario B.  Distance $y$ with initial parameter $K^{(1)}$ (black line), optimal parameter $K^{*}$ (red line) and mean parameter $\hat{K}$ (yellow line).}\label{fig:dist_RV_comparison}
	\end{figure}

	Figure~\ref{fig:K_RV} shows the evolution of the parameter vector $K^{(i)}$ during the learning. The resulting final mean parameter vector is equal to $\hat{K}=\left[1.22,\,5.41,\,0.72,\,5.29,\,0.40\right]^T$. Fig.~\ref{fig:dist_RV_comparison} reports a comparison among the trajectories $y$ obtained by the control laws \eqref{eq:controller} with initial parameter $K^{(1)}$, optimal parameter $K^{*}$ and mean parameter $\hat{K}$, respectively,  for one of the considered initial conditions. It can be seen that the mean controller leads to a reduction of the convergence time which is comparable to the one obtained by the optimal parameters $K^*$ for that initial condition. The statistics of this experimental campaign are summarized in Table~\ref{tab:RVstat}, indicating that the two cost reductions achieved are comparable. This finding suggests that, despite the large variability achieved by the learned parameters (see  Fig.~\ref{fig:K_RV}), utilizing the mean parameter vector $\hat{K}$ in the controller \eqref{eq:controller} for terminal rendezvous maneuvers is a reasonable option, allowing for very good performance without demanding on-board computations. 
{Conversely, it is evident that a coarse parameter tuning entails a severe performance degradation.} Therefore, the use of pre-computed mean parameter vectors can be an effective strategy for achieving reliable performance in scenarios where the initial conditions are not known \textit{a priori} and the computational resources are limited.
	
	
	\begin{table}[h]
		\centering 
		\renewcommand{\arraystretch}{1.3}
		\begin{tabular}{|l|cc|}
			\hline
			\multirow{2}{*}{} & \multicolumn{2}{c|}{Total cost reduction (\%)} \\ \cline{2-3} 
			& \multicolumn{1}{c|}{$K^{*}$} & $\hat{K}$ \\ \hline
			Average & \multicolumn{1}{c|}{81.3} & 78.4 \\
			Minimum & \multicolumn{1}{c|}{72.6} & 68.9 \\
			Maximum & \multicolumn{1}{c|}{87.2} & 85.9 \\ \hline
		\end{tabular}
		\caption{Performance results for Scenario B with random initial conditions.}\label{tab:RVstat}
	\end{table}

	\section{Conclusions}
	\label{sec:conc}
	Optimization of performance measures in orbital tracking is a challenging task due to the complexity of the dynamic models and the necessity to guarantee fundamental requirements such as stability, robustness and constraint satisfaction. This work has shown that machine learning techniques can be successfully employed to tune the parameters of a family of stabilizing controllers for orbital tracking, in order to optimize a cost function accounting for both settling time and fuel consumption. In perspective, this approach can be also useful to analyze the sensitivity of the performance metrics with respect to the control parameters. Besides the considered augmented random search algorithm, future investigation may concern other learning approaches (e.g., policy gradient) and address the inclusion of state/input constraints or parametric uncertainties in the optimization problem.

	\bibliographystyle{IEEEtran}
	\balance
	\bibliography{rl,ml4gnc}

\begin{thebibliography}{10}
\providecommand{\url}[1]{#1}
\csname url@samestyle\endcsname
\providecommand{\newblock}{\relax}
\providecommand{\bibinfo}[2]{#2}
\providecommand{\BIBentrySTDinterwordspacing}{\spaceskip=0pt\relax}
\providecommand{\BIBentryALTinterwordstretchfactor}{4}
\providecommand{\BIBentryALTinterwordspacing}{\spaceskip=\fontdimen2\font plus
\BIBentryALTinterwordstretchfactor\fontdimen3\font minus
  \fontdimen4\font\relax}
\providecommand{\BIBforeignlanguage}[2]{{%
\expandafter\ifx\csname l@#1\endcsname\relax
\typeout{** WARNING: IEEEtran.bst: No hyphenation pattern has been}%
\typeout{** loaded for the language `#1'. Using the pattern for}%
\typeout{** the default language instead.}%
\else
\language=\csname l@#1\endcsname
\fi
#2}}
\providecommand{\BIBdecl}{\relax}
\BIBdecl

\bibitem{AWbook13}
K.~J. {\AA}str{\"o}m and B.~Wittenmark, \emph{Adaptive control}.\hskip 1em plus
  0.5em minus 0.4em\relax Courier Corporation, 2013.

\bibitem{KKKbook95}
M.~Krstic, P.~V. Kokotovic, and I.~Kanellakopoulos, \emph{Nonlinear and
  adaptive control design}.\hskip 1em plus 0.5em minus 0.4em\relax John Wiley
  \& Sons, Inc., 1995.

\bibitem{MDB92}
K.~L. Moore, M.~Dahleh, and S.~Bhattacharyya, ``Iterative learning control: A
  survey and new results,'' \emph{Journal of Robotic Systems}, vol.~9, no.~5,
  pp. 563--594, 1992.

\bibitem{BTA06}
D.~A. Bristow, M.~Tharayil, and A.~G. Alleyne, ``A survey of iterative learning
  control,'' \emph{IEEE Control Systems Magazine}, vol.~26, no.~3, pp. 96--114,
  2006.

\bibitem{CLS02}
M.~C. Campi, A.~Lecchini, and S.~M. Savaresi, ``Virtual reference feedback
  tuning: a direct method for the design of feedback controllers,''
  \emph{Automatica}, vol.~38, no.~8, pp. 1337--1346, 2002.

\bibitem{FN15}
L.~Fagiano and C.~Novara, ``Learning a nonlinear controller from data: Theory,
  computation, and experimental results,'' \emph{IEEE Transactions on Automatic
  Control}, vol.~61, no.~7, pp. 1854--1868, 2015.

\bibitem{Ber19book}
D.~Bertsekas, \emph{Reinforcement learning and optimal control}.\hskip 1em plus
  0.5em minus 0.4em\relax Athena Scientific, 2019.

\bibitem{Rec19}
B.~Recht, ``A tour of reinforcement learning: The view from continuous
  control,'' \emph{Annual Review of Control, Robotics, and Autonomous Systems},
  vol.~2, pp. 253--279, 2019.

\bibitem{MKSR15}
V.~Mnih, K.~Kavukcuoglu, D.~Silver, A.~A. Rusu, J.~Veness, M.~G. Bellemare,
  A.~Graves, M.~Riedmiller, A.~K. Fidjeland, G.~Ostrovski \emph{et~al.},
  ``Human-level control through deep reinforcement learning,'' \emph{Nature},
  vol. 518, no. 7540, pp. 529--533, 2015.

\bibitem{SHMG16}
D.~Silver, A.~Huang, C.~J. Maddison, A.~Guez, L.~Sifre, G.~Van Den~Driessche,
  J.~Schrittwieser, I.~Antonoglou, V.~Panneershelvam, M.~Lanctot \emph{et~al.},
  ``Mastering the game of go with deep neural networks and tree search,''
  \emph{Nature}, vol. 529, no. 7587, pp. 484--489, 2016.

\bibitem{LHPH15}
T.~P. Lillicrap, J.~J. Hunt, A.~Pritzel, N.~Heess, T.~Erez, Y.~Tassa,
  D.~Silver, and D.~Wierstra, ``Continuous control with deep reinforcement
  learning,'' \emph{arXiv preprint arXiv:1509.02971}, 2015.

\bibitem{KIPI18}
D.~Kalashnikov, A.~Irpan, P.~Pastor, J.~Ibarz, A.~Herzog, E.~Jang, D.~Quillen,
  E.~Holly, M.~Kalakrishnan, V.~Vanhoucke \emph{et~al.}, ``Scalable deep
  reinforcement learning for vision-based robotic manipulation,'' in
  \emph{Conference on Robot Learning}.\hskip 1em plus 0.5em minus 0.4em\relax
  PMLR, 2018, pp. 651--673.

\bibitem{ABCJ20}
M.~Andrychowicz, B.~Baker, M.~Chociej, R.~Jozefowicz, B.~McGrew, J.~Pachocki,
  A.~Petron, M.~Plappert, G.~Powell, A.~Ray \emph{et~al.}, ``Learning dexterous
  in-hand manipulation,'' \emph{The International Journal of Robotics
  Research}, vol.~39, no.~1, pp. 3--20, 2020.

\bibitem{GBKD21}
L.~C. Garaffa, M.~Basso, A.~A. Konzen, and E.~P. de~Freitas, ``Reinforcement
  learning for mobile robotics exploration: A survey,'' \emph{IEEE Transactions
  on Neural Networks and Learning Systems}, 2021.

\bibitem{PBYV17}
X.~B. Peng, G.~Berseth, K.~Yin, and M.~Van De~Panne, ``Deeploco: Dynamic
  locomotion skills using hierarchical deep reinforcement learning,'' \emph{ACM
  Transactions on Graphics (TOG)}, vol.~36, no.~4, pp. 1--13, 2017.

\bibitem{ZZQ19}
Z.~Zhang, D.~Zhang, and R.~C. Qiu, ``Deep reinforcement learning for power
  system applications: An overview,'' \emph{CSEE Journal of Power and Energy
  Systems}, vol.~6, no.~1, pp. 213--225, 2019.

\bibitem{izzo2019survey}
D.~Izzo, M.~M{\"a}rtens, and B.~Pan, ``A survey on artificial intelligence
  trends in spacecraft guidance dynamics and control,'' \emph{Astrodynamics},
  pp. 1--13, 2019.

\bibitem{shirobokov2021}
M.~Shirobokov, S.~Trofimov, and M.~Ovchinnikov, ``Survey of machine learning
  techniques in spacecraft control design,'' \emph{Acta Astronautica}, vol.
  186, pp. 87--97, 2021.

\bibitem{Xia2016}
K.~Xia and W.~Huo, ``Robust adaptive backstepping neural networks control for
  spacecraft rendezvous and docking with uncertainties,'' \emph{Nonlinear
  Dynamics}, vol.~84, pp. 1683--1695, 2016.

\bibitem{Ueda2019}
S.~Ueda and A.~Noumi, ``Precise rendezvous guidance in low earth orbit via
  machine learning,'' in \emph{Proceedings of SICE International Symposium on
  Control Systems, SICE ISCS}, 2019.

\bibitem{Hongjue2020}
P.~L. Hongjue~Li, Yunfeng~Dong, ``Real-time optimal approach and capture of
  {ENVISAT} based on neural networks,'' \emph{International Journal of
  Aerospace Engineering}, vol. 2020, pp. 1--17, 2020.

\bibitem{Gaudet2018}
B.~Gaudet, R.~Linares, and R.~Furfaro, ``{Spacecraft rendezvous guidance in
  cluttered environments via artificial potential functions and reinforcement
  learning},'' in \emph{Advances in the Astronautical Sciences}, vol. 167,
  2018, pp. 813--828.

\bibitem{Wang2020}
X.~Wang, G.~Wang, Y.~Chen, and Y.~Xie, ``Autonomous rendezvous guidance via
  deep reinforcement learning,'' in \emph{2020 Chinese Control And Decision
  Conference (CCDC)}.\hskip 1em plus 0.5em minus 0.4em\relax IEEE, 2020, pp.
  1848--1853.

\bibitem{hovell2020deep}
K.~Hovell and S.~Ulrich, ``On deep reinforcement learning for spacecraft
  guidance,'' in \emph{AIAA Scitech 2020 Forum}, 2020.

\bibitem{BTSK17}
F.~Berkenkamp, M.~Turchetta, A.~Schoellig, and A.~Krause, ``Safe model-based
  reinforcement learning with stability guarantees,'' \emph{Advances in neural
  information processing systems}, vol.~30, 2017.

\bibitem{YSA21}
H.~Yin, P.~Seiler, and M.~Arcak, ``Stability analysis using quadratic
  constraints for systems with neural network controllers,'' \emph{IEEE
  Transactions on Automatic Control}, vol.~67, no.~4, pp. 1980--1987, 2021.

\bibitem{WBRM21}
R.~Wang, N.~Barbara, M.~Revay, and I.~R. Manchester, ``Learning over all
  stabilizing nonlinear controllers for a partially-observed linear system,''
  \emph{arXiv preprint arXiv:2112.04219}, 2021.

\bibitem{newton2022}
M.~Newton and A.~Papachristodoulou, ``Stability of non-linear neural feedback
  loops using sum of squares,'' in \emph{2022 IEEE 61st Conference on Decision
  and Control (CDC)}.\hskip 1em plus 0.5em minus 0.4em\relax IEEE, 2022, pp.
  6000--6005.

\bibitem{RMT11}
J.~W. Roberts, I.~R. Manchester, and R.~Tedrake, ``Feedback controller
  parameterizations for reinforcement learning,'' in \emph{2011 IEEE Symposium
  on Adaptive Dynamic Programming and Reinforcement Learning (ADPRL)}.\hskip
  1em plus 0.5em minus 0.4em\relax IEEE, 2011, pp. 310--317.

\bibitem{Wil92}
R.~J. Williams, ``Simple statistical gradient-following algorithms for
  connectionist reinforcement learning,'' \emph{Machine learning}, vol.~8,
  no.~3, pp. 229--256, 1992.

\bibitem{FB17}
S.~R. Friedrich and M.~Buss, ``A robust stability approach to robot
  reinforcement learning based on a parameterization of stabilizing
  controllers,'' in \emph{2017 IEEE International Conference on Robotics and
  Automation (ICRA)}.\hskip 1em plus 0.5em minus 0.4em\relax IEEE, 2017, pp.
  3365--3372.

\bibitem{LSLF20}
N.~P. Lawrence, G.~E. Stewart, P.~D. Loewen, M.~G. Forbes, J.~U. Backstrom, and
  R.~B. Gopaluni, ``Reinforcement learning based design of linear fixed
  structure controllers,'' \emph{IFAC-PapersOnLine}, vol.~53, no.~2, pp.
  230--235, 2020.

\bibitem{LFLM22}
N.~P. Lawrence, M.~G. Forbes, P.~D. Loewen, D.~G. McClement, J.~U.
  Backstr{\"o}m, and R.~B. Gopaluni, ``Deep reinforcement learning with shallow
  controllers: An experimental application to {PID} tuning,'' \emph{Control
  Engineering Practice}, vol. 121, p. 105046, 2022.

\bibitem{HST20}
T.~Holicki, C.~W. Scherer, and S.~Trimpe, ``Controller design via experimental
  exploration with robustness guarantees,'' \emph{IEEE Control Systems
  Letters}, vol.~5, no.~2, pp. 641--646, 2020.

\bibitem{leomanni2017class}
M.~Leomanni, G.~Bianchini, A.~Garulli, and A.~Giannitrapani, ``A class of
  globally stabilizing feedback controllers for the orbital rendezvous
  problem,'' \emph{International Journal of Robust and Nonlinear Control},
  vol.~27, no.~18, pp. 4607--4621, 2017.

\bibitem{walker1985set}
M.~J.~H. Walker, B.~Ireland, and J.~Owens, ``A set of modified equinoctial
  orbit elements,'' \emph{Celestial mechanics}, vol.~36, no.~4, pp. 409--419,
  1985.

\bibitem{battin1999introduction}
R.~H. Battin, \emph{An Introduction to the Mathematics and Methods of
  Astrodynamics}.\hskip 1em plus 0.5em minus 0.4em\relax Reston: {AIAA}, 1999.

\bibitem{MGR18}
H.~Mania, A.~Guy, and B.~Recht, ``Simple random search of static linear
  policies is competitive for reinforcement learning,'' \emph{Advances in
  Neural Information Processing Systems}, vol.~31, 2018.

\end{thebibliography}
\end{document}